\documentclass[prd, aps, showpacs]{revtex4}
\usepackage{latexsym, graphicx}

\def\be{\begin{equation}}
\def\te{\end{equation}}
\def\bea{\begin{eqnarray}}

\def\tea{\end{eqnarray}}
\begin{document}

\title{Gauge-invariant Effective Action for the Dynamics of Bose-Einstein
condensates with a fixed number of atoms}
\author{Esteban Calzetta$^{1}$ and B. L. Hu$^{2}$ \\
%EndAName
$^{1}${\small Departamento de Fisica, Facultad de Ciencias Exactas y
Naturales,}\\
{\small Universidad de Buenos Aires- Ciudad Universitaria, 1428 Buenos
Aires, Argentina}\\
$^{2}${\small Department of Physics, University of Maryland, College Park,
MD 20742}}
\date{\today }

\begin{abstract}
In this paper we present a particle-number-conserving (PNC) functional
formalism to describe the dynamics of a cold bosonic gas. Treating the total
number of particles as a constraint, whereby the phase invariance of the
theory becomes local in time, we study this U(1) gauge theory using DeWitt's
"gauge invariant effective action" techniques. Our functional formulation
and earlier PNC proposals are shown to yield equivalent results to
next-to-leading order in an expansion in the inverse powers of the total
number of particles. In this more general framework we also show that
earlier PNC proposals can be seen as different gauge (and gauge fixing
condition) choices within the same physical theory.
\end{abstract}

\maketitle

\section{Introduction}

Experimental successes in the making of Bose-Einstein condensates (BEC)
ushered in a new epoch in atomic and optical physics where precision control
of many-atom systems becomes possible. From this we see the rapid
development of new tools and techniques where one can construct such systems
designed for probing deeper physical issues as well as for more practical
purposes (such as quantum information processing QIP). With these advances
more sophisticated theoretical methods need be introduced to describe the
quantum field and nonequilibrium properties of such systems, addressing the
issues related to quantum coherence, correlations, fluctuations and noise.
The demand here is to treat the dynamics of strongly correlated quantum
systems often in the non-Markovian regimes while respecting the full quantum
coherence of the system, meeting the constraints from the set-up and
incorporating the effect of noise and fluctuations.

Technically speaking, because any consistent description of quantum
fluctuations around a BEC \cite{PS02} is gapless \cite{HM65,GR96},
individual Feynman graphs contributing to quantities of interest may be
large. An accurate description of the full quantum dynamics requires
resummation of whole classes of graphs. A familiar example is the hard
thermal loop techniques developed to describe hot quark-gluon plasmas \cite
{LeBellac}. These nonperturbative approaches are most easily implemented at
the level of the so-called closed time-path effective action (CTPEA)\cite
{ctp,StoofCTP,LundhRammer}, closely related to the Feynman - Vernon
influence functional for quantum open systems \cite{if}.

The CTPEA allows for two complementary lines of development. If the focus of
interest is the evolution of quantum fluctuations and their correlations,
then the CTPEA is best combined with the so-called 2 particle irreducible
(or Cornwall - Jackiw -Tomboulis) \cite{CJT} effective action to yield
dynamical equations of motion for the correlations themselves \cite{CH88}.
This is in fact the method of choice for dealing with quantum fields in far
from equilibrium situations \cite{Bergesb}. We should stress that the
formulation of the BEC dynamics as a hierarchy of correlations is well known 
\cite{BurKoh}; what sets the EA approach apart is its efficiency with
respect to the implementation of nonperturbative resummation of diagrams 
\cite{ByHFB,Ana2,GBSS05}.

If the focus of interest is the condensate itself, then the CTPEA may be
used to derive a stochastic Gross - Pitaievsky (GP) equation, whereby
quantum expectation values of the full theory may be retrieved as averages
over the noise. In gravitation and cosmology, a similar problem has led to
the derivation of the so-called Einstein-Langevin equation as the
centerpiece of the stochastic semiclassical gravity program \cite{SSG}.
Different approaches have been proposed \cite{GarAngDav} to tackle the
interaction of BEC with noncondensate atoms. While these approaches can be
shown to be equivalent in the appropriate limits, derivation from the CTPEA
enables one to keep track of the non-Markovian nature of dissipation and
(colored) noise in a self-consistent manner. This formulation may also be
applied to the treatment of decoherence of BEC in microtraps \cite{HenGar},
a problem similar to environment-induced decoherence in the context of
structure formation via the influence functional and CTP techniques \cite
{CH95,LMR03}.

Functional methods have been introduced for the description of BEC dynamics
(see, e.g., the lectures of Stoof \cite{StoofCTP}). The advantages of a
field theory approach to many-body systems are well-known: economy in the
formulation, flexibility in the use of self-consistency to improve on
perturbation theory and the explicit enforcement of conservation laws. There
is a vast literature on this subject based on the so - called ``symmetry
breaking'' approach: the theory has a global U(1) symmetry whose breakdown
signals condensation. In this approach particle number is conserved but not
fixed: there are quantum fluctuations in both the number of particles in the
condensate and the total number of particles in the system, only the mean
numbers are fixed.

Though convenient this approach contradicts with the actual experimental
situation, where the total number of particles is fixed or, if there is loss
of atoms, this needs be accounted for in real physical terms. Moreover, in
typical experiments the total number of particles (several thousands) is not
large enough to make the difference negligible. With precision
experimentation, especially for QIP purposes, one should seek a more
accurate method. This is the primary motivation for the present work.

To the best of our knowledge, the first proposal for handling a BEC with a
fixed number of particles has been put forward by Arnowitt and Girardeau 
\cite{GA59}, leading to the development of the so-called particle number
conserving (PNC) method \cite{GAR97,CD98,G98,GJDCZ00,MOR03,DS03}. The goal
of this paper is to formulate a functional PNC approach, and to show the
equivalence of the functional and those earlier proposals to next to leading
order (NLO) in an expansion in inverse powers of the total number of
particles. Let us first see what the problem entails with some technical
help.

\subsection{Field theory approach to BECs}

In recent years, field theory techniques have been systematically applied to
the dynamics of BECs \cite{PS02,AND04}. The most common approach is based on
second - quantization formalism for a many-atom system. The field operator $%
\Psi \left( x,t\right) $ which removes an atom at the location $x$ at times $%
t$ obeys the canonical commutation relations:

\begin{equation}
\left[ \Psi \left( x,t\right) ,\Psi \left( y,t\right) \right] =0
\end{equation}
\begin{equation}
\left[ \Psi \left( x,t\right) ,\Psi \left( y,t\right) \right] =\delta \left(
x-y\right)  \label{etcr}
\end{equation}
The dynamics of this field is given by the Heisenberg equations of motion

\begin{equation}
-i\hbar \frac{\partial }{\partial t}\Psi =\left[ \mathbf{H},\Psi \right]
\label{Hei1}
\end{equation}
where $\mathbf{H}$ is the Hamiltonian

\begin{equation}
\mathbf{H}=\int d^{d}x\;\left\{ \Psi H\Psi +\frac{U}{2}\left( \Psi \Psi
\right) ^{2}\right\}  \label{nbodyh}
\end{equation}

\begin{equation}
H\Psi =-\frac{\hbar ^{2}}{2M}\nabla ^{2}\Psi +V_{trap}\left( x\right) \Psi
\label{sparth}
\end{equation}
and $V_{trap}\left( x\right) $ denotes a confining trap potential. Making
the commutator in Eq. (\ref{Hei1}) explicit we obtain

\begin{equation}
i\hbar \frac{\partial }{\partial t}\Psi =H\Psi +U\Psi ^{\dagger }\Psi ^{2}
\label{Hei2}
\end{equation}

The total particle number operator commutes with the Hamiltonian and is
therefore conserved. The theory is invariant under a global phase change of
the field operator 
\begin{equation}
\Psi \rightarrow e^{i\theta }\Psi ,\qquad \Psi \rightarrow e^{-i\theta }\Psi
\label{global}
\end{equation}

\subsection{Problems with the symmetry breaking approach}

In the symmetry breaking (SB) approach to BECs, condensation is signaled by
the spontaneous breakdown of phase invariance Eq. (\ref{global}), whereby $%
\Psi $ develops a nonzero expectation value $\Phi $ (c-number). We can
therefore employ a background field separation for $\Psi $ \cite
{CL95,NO98,PS02}

\begin{equation}
\Psi =\Phi _{SB}+{\Psi }_{q}  \label{mfs}
\end{equation}
where {$\Psi _{q}$} ($q$- number) is the field operator corresponding to
quantum fluctuations. Various approaches differ in how to handle the
dynamics of these two constituents.

For later reference, we point out that a priori {$\Psi _{q}$} is not
necessarily orthogonal to $\Phi _{SB};$ also $\Phi _{SB}$ is not necessarily
\ simply proportional to the condensate wave function as defined below.

A common feature of these approaches is that the total particle number 
\begin{equation}
\mathbf{N}=\int d^{d}x\;\Psi ^{\dagger }\Psi  \label{num}
\end{equation}
is not fixed. For example, let us assume that the condensate is confined
within a homogeneous box of volume $V$, condensation occurring in the lowest
(translation invariant) mode. Let $a_{\vec{k}}$ be the operator that
destroys an atom in the $\vec{k}$ mode. Then we may approximate (see more
careful discussion below)

\begin{equation}
\Psi _{q}\left( x,t\right) =\sum_{\vec{k}\neq 0}\frac{e^{i\vec{k}\vec{x}}}{%
\sqrt{V}}a_{\vec{k}}
\end{equation}

Even if we treat $\Psi _{q}$ as a linear perturbation on the condensate, the
Hamiltonian is not diagonal on the $a_{\vec{k}}.$ To diagonalize it, we must
introduce phonon destruction operators $b_{\vec{k}}$ and perform a Bogolubov
transformation

\begin{equation}
a_{\vec{k}}=\alpha _{k}b_{\vec{k}}+\beta _{k}b_{-\vec{k}}^{\dagger }
\end{equation}

At zero temperature, the state is the phonon vacuum, $b_{\vec{k}}\left|
0\right\rangle =0$ for all $\vec{k}\neq 0$. We find

\begin{equation}
\left\langle \mathbf{N}\right\rangle =\int d^{d}x\;\left\langle \Psi
^{\dagger }\Psi \right\rangle =V\left[ \left| \Phi _{SB}\right| ^{2}+\tilde{n%
}\right]
\end{equation}
where 
\begin{equation}
\tilde{n}=\left\langle \Psi _{q}^{\dagger }\Psi _{q}\right\rangle =\frac{1}{V%
}\sum_{\vec{k}\neq 0}\left\langle a_{\vec{k}}^{\dagger }a_{\vec{k}%
}\right\rangle =\frac{1}{V}\sum_{\vec{k}\neq 0}\left| \beta _{k}\right| ^{2}
\end{equation}
but

\begin{equation}
\left\langle \mathbf{N}^{2}\right\rangle =V^{2}\left[ \left( \left| \Phi
_{SB}\right| ^{2}\right) ^{2}+\left| \Phi _{SB}\right| ^{2}\left( 4\tilde{n}+%
\frac{1}{V}\right) +\Phi _{SB}^{*2}\tilde{m}+\Phi _{SB}^{2}\tilde{m}%
^{*}+...\right]
\end{equation}
where 
\begin{equation}
\tilde{m}=\left\langle \Psi _{q}^{2}\right\rangle =\frac{1}{V}\sum_{\vec{k}%
\neq 0}\left\langle a_{-\vec{k}}a_{\vec{k}}\right\rangle =\frac{1}{V}\sum_{%
\vec{k}\neq 0}\alpha _{k}\beta _{k}
\end{equation}

The Bogoliubov coefficients $\alpha _{k}$ and $\beta _{k}$ cannot be equal,
because the canonical (bosonic) commutation relations imply $\left| \alpha
_{k}\right| ^{2}-\left| \beta _{k}\right| ^{2}=1,$ and so also $\tilde{m}%
\neq \tilde{n}.$ We conclude that necessarily $\left\langle \mathbf{N}%
^{2}\right\rangle \neq \left\langle \mathbf{N}\right\rangle ^{2}$ in the
symmetry breaking approach, signaling the presence of particle number
fluctuations.

\subsection{Particle number conserving functional approach: Global versus
local gauge}

The Heisenberg equation of motion Eq. (\ref{Hei2}) is also the classical
equation of motion derived from the action

\begin{equation}
S=\int d^{d+1}x\;i\hbar \Psi ^{\ast }\frac{\partial }{\partial t}\Psi -\int
dt\;\mathbf{H}  \label{action}
\end{equation}

The quantum theory of the BEC may be regarded as the quantization of the
nonrelativistic classical field theory defined by the action functional Eq. (%
\ref{action}), where the canonical variables are $\Psi \left( x,t\right) $
and its conjugate momentum $i\hbar \Psi ^{*}.$ This theory conserves
particle number Eq. (\ref{num}), and we are interested in the case in which
particle number takes on a definite value $N$. We may reinforce this point
by adding a constraint on the theory. This is achieved by introducing a
Lagrange multiplier $\mu _{q}\left( t\right) ,$ thereby writing the action as

\begin{equation}
S=\int d^{d+1}x\;\left\{ i\hbar \Psi ^{*}\frac{\partial }{\partial t}\Psi
+\hbar \mu _{q}\left( t\right) \left[ \Psi ^{*}\Psi -\frac{N}{V}\right]
\right\} -\int dt\;\mathbf{H}  \label{newaction}
\end{equation}

The original action Eq. (\ref{action}) is invariant under a global
transformation Eq. (\ref{global}) but the new action Eq. (\ref{newaction})
is invariant under the local (in time) transformations (a familiar theory
with local U(1) gauge symmetry is electromagnetism)

\begin{equation}
\Psi \rightarrow e^{i\theta \left( t\right) }\Psi ,\qquad \Psi ^{\dagger
}\rightarrow e^{-i\theta \left( t\right) }\Psi ^{\dagger },\qquad \mu
_{q}\rightarrow \mu _{q}+\frac{d\theta }{dt}  \label{local}
\end{equation}
provided $\theta $ vanishes both at the initial and final times (when $%
\theta $ is infinitesimal, these are just canonical transformations
generated by the constraint). Therefore it must be quantized using the
methods developed for gauge theories, such as the Fadeev-Popov method (See
Appendix B).

In short, the action functionals Eq. (\ref{action}) and (\ref{newaction})
lead to the same classical theories, but to inequivalent quantization
schemes. The assertion is that Eq. (\ref{newaction}) gives a more complete
description of the fundamental theory, as it shows explicitly the gauge
dependence while the result of the quantum theory is gauge-invariant. As it
has been shown by DeWitt and others, it is possible to introduce an action
functional whose variation yields the equations of motion for the
expectation values of the fields and preserves the gauge symmetry. The
explicit construction of this so-called ''gauge invariant effective action''
(GIEA) \cite{GIEA} for the description of BEC dynamics is the subject of
this paper.

\subsection{Specific goals and what are attained}

In this paper we shall construct the gauge-invariant effective action (GIEA)
for the dynamics of BECs with a fixed total particle number. The variation
of this GIEA yields the equations of motion for the background fields.

Since physical description of results depends on the choice of gauge many
existing approaches which work within a specific gauge produce results which
cannot easily be compared with each other. In the GIEA approach we can place
them in the same framework and be able to compare the similarities and
difference in the physics. For a range of existing theories with particle
number conserving constraints \cite{GA59,GAR97,CD98,GJDCZ00,MOR03} we show
that they are indeed within the same physical theory. Most importantly, the
gauge-invariant formalism provides the necessary foundation to carry on
these approaches beyond the next to leading order approximation in a
consistent way. It is in this more ambitious program that the power of the
gauge invariant approach becomes decisive, not optional.

Our goal in this paper is to construct the basic structure for such a
program, putting premium emphasis on the gauge-invariance aspect but
limiting our reach only to the ``one-particle irreducible'' (1PI) level \cite
{PS95}, as opposed to a 2PI or $\Phi $- derivable approach \cite{CJT}. The
extension to the $\Phi $- derivable formulation (or even further, to the nPI
effective action \cite{nPI}) is necessary to match the accuracy demanded by
present day experiments. A 2PI non-gauge theory description of the
nonequilibrium dynamics of BECs is presented in \cite{ByHFB,Ana2,GBSS05};
the 2PI approach in gauge theories at large is discussed in \cite{Peyresq8}.

Another restriction is that we shall work with the IN-OUT formulation \cite
{PS95}, as opposed to the closed - time path (CTP) or Schwinger - Keldysh
formulation \cite{ctp}. Beyond the next-to-leading order a truly dynamical
theory requires a CTP formulation which yields real and causal equations of
motion \cite{CH87}. Formally, the CTP dynamical equations may be derived
from those in this paper by considering the time variable to be defined on a
closed time path ranging from the distant past to the far future and
doubling back to the distant past. For further discussion we refer the
reader to Refs. \cite{ByHFB,Ana2}

This paper is organized as follows. In Section II we develop a PNC\ approach
in canonical terms, which will be later used as a template for the
functional approach. Our presentation is close to \cite{CD98}, with some
minor technical differences (explained in Appendix A) 
% we compare the formalism in Section II to those of
%Girardeau and Arnowitt \cite {GA59,G98}, Gardiner \cite{GAR97} and Castin and Dum \cite{CD98}.
The essence of the paper is in Section III, where we construct the GIEA for
the description of BECs with a fixed number of particles. We derive the
equations of motion in a large $N$ expansion, where $N$ is the number of
particles, and show their equivalence to those derived by canonical methods
in Section II, up to NLO. (Observe that in the literature there is also a
``large $\mathcal{N}"$ expansion, where $\mathcal{N}$ is the number of
fields \cite{GBSS05,AND04}. In our case, $\mathcal{N}=2$ from $\Psi $ and $%
\Psi ^{\dagger }$. These expansions should not be confused with each other.)

In Section IV we conclude with some brief remarks. Appendix A compares our
presentation of the PNC formalism with those of Girardeau and Arnowitt \cite
{GA59,G98}, Gardiner \cite{GAR97} and Castin and Dum \cite{CD98}, and
Appendix B gives a review of gauge field theory quantization and the theory
of the GIEA . None of the material in both Appendices is new, they are added
for completeness and easy reference.

\section{The particle number conserving formalism}

The symmetry-breaking approach described above has the disturbing feature
that, strictly speaking, symmetry breaking only occurs in the thermodynamic
limit. We have therefore a formalism that assumes the number of particles is
essentially infinite. Most actual experiments deal with situations where
particle number is bounded and actually not so large (from a few hundred to
a few thousand atoms). Under this circumstance a condensate as described
above simply cannot happen.

In this Section we shall describe an alternative formulation which is
designed to deal with gases at a fixed particle number. We shall call this
formulation the particle number conserving formalism, PNC for short. It was
first propossed by Arnowitt and Girardeau \cite{GA59}, and it has been
extended and improved by many other people over the last forty years \cite
{GAR97,G98}. Our own presentation follows Castin and Dum \cite
{CD98,MOR03,DS03}, with some minor technical differences which are discussed
in Appendix A.

Let us begin by discussing how is it possible to speak of a BEC in a
situation where there is no symmetry breaking.

\subsection{The one-body density matrix and long range coherence}

We consider as above a second - quantized Bose field $\Psi $. The state of
the many-body system is an eigenstate of total particle number operator Eq. (%
\ref{num}). There is no particle exchange with the environment.

In this case of a finite system, there is no symmetry breaking. The symmetry
broken state is essentially a coherent state and thus a coherent
superposition of states with arbitrarily large total particle number.
Nevertheless, there are situations where there is long range coherence
across the system, thus capturing the essential feature of the condensed
states. Sometimes these situations are referred to as quasi-condensates, but
we shall not make this distinction, and call them BECs as their symmetry -
broken siblings.

To characterize the BEC state, let us introduce the one-body density matrix 
\cite{PO56}

\begin{equation}
\sigma \left( x,y,t\right) =\left\langle \Psi ^{\dagger }\left( x,t\right)
\Psi \left( y,t\right) \right\rangle
\end{equation}
Long range coherence appears when $\sigma $ fails to decay as $x$ and $y$
are taken apart.

Observe that $\sigma $ is Hermitian and nonnegative, in the sense that for
any function $f$

\begin{equation}
\int d^{d}xd^{d}y\;f^{*}\left( x\right) \sigma \left( x,y,t\right) f\left(
y\right) \succeq 0
\end{equation}
Therefore it admits a basis of eigenfunctions

\begin{equation}
\int d^{d}x\;\sigma \left( x,y,t\right) \phi _{\alpha }\left( y,t\right)
=n_{\alpha }\phi _{\alpha }\left( x,t\right)
\end{equation}
where the eigenvalues $n_{\alpha }$ are real and nonnegative. We assume the $%
\phi _{\alpha }$ are normalized

\begin{equation}
\left( \phi _{\alpha },\phi _{\beta }\right) =\delta _{\alpha \beta }
\end{equation}

\begin{equation}
\left( f,g\right) =\int d^{d}x\;f^{*}g
\end{equation}
and complete

\begin{equation}
\sum_{\alpha }\phi _{\alpha }^{*}\left( x,t\right) \phi _{\alpha }\left(
y,t\right) =\delta \left( x-y\right)
\end{equation}

The field operator may be expanded in this basis

\begin{equation}
\Psi \left( x,t\right) =\sum_{\alpha }a_{\alpha }\left( t\right) \phi
_{\alpha }\left( x,t\right)
\end{equation}
The Bose commutation relations imply

\begin{equation}
\left[ a_{\alpha }\left( t\right) ,a_{\beta }^{\dagger }\left( t\right)
\right] =\delta _{\alpha \beta }
\end{equation}
The $a_{\alpha }\left( t\right) $ are operators which, at time $t$, destroy
a particle in the one-particle state $\alpha $ whose wavefunction is $\phi
_{\alpha }\left( x,t\right) .$ From the definition of $\sigma $ we find

\begin{equation}
\left\langle a_{\alpha }^{\dagger }\left( t\right) a_{\beta }\left( t\right)
\right\rangle =n_{\alpha }\left( t\right) \delta _{\alpha \beta }
\end{equation}
Therefore the eigenvalues $n_{\alpha }\left( t\right) $ are the mean number
of particles in the one-body state $\alpha $ at time $t.$ We also have the
strong identity

\begin{equation}
N=\sum_{\alpha }a_{\alpha }^{\dagger }\left( t\right) a_{\alpha }\left(
t\right)
\end{equation}

Condensation occurs when one of the $n_{\alpha },$ say $\alpha =0,$ becomes
comparable with $N$ itself. Then we have, for large separations

\begin{equation}
\sigma \left( x,y,t\right) \sim n_{0}\phi _{0}^{*}\left( x,t\right) \phi
_{0}\left( y,t\right)
\end{equation}
which displays long range coherence, as expected. Here $\phi _{0}\left(
x,t\right) $ is the condensate wave function. We must stress that this is
the fundamental definition; $\phi _{0}\left( x,t\right) $ is not necessarily
identical to the mean field $\Phi _{SB}$ introduced in the symmetry -
breaking approach (for one, $\Phi _{SB}$ is not normalized).

\subsection{The dynamics of the condensate wave function}

We shall now discuss the dynamics of the condensate wave function $\phi
_{0}\left( x,t\right) $ and the condensate occupation number $N_{0}$ (we
switch to a capital $N$ to emphasize its macroscopic character). We envisage
a situation in which $N$ is finite but large, and will seek equations of
motion as an expansion in inverse powers of $N.$ In preparation for this, it
is convenient to scale the interaction term, writing $U=u/N.$

The idea is to exploit the fact that $a_{0}$ is ``large'' and the other
destruction operators are not. Let us define new operators $\lambda _{\alpha
},$ $\alpha \neq 0,$ through

\begin{equation}
a_{\alpha }\equiv \frac{a_{0}}{\sqrt{N}}\lambda _{\alpha }  \label{newop}
\end{equation}
The $\lambda _{\alpha }$ do not commute with $a_{0}^{\dagger }$

\begin{equation}
a_{0}\left[ \lambda _{\alpha },a_{0}^{\dagger }\right] +\lambda _{\alpha }=0
\end{equation}
but they commute with $N$. Let us define the non-condensate field

\begin{equation}
\Lambda \left( x,t\right) =\sum_{\beta \neq 0}\lambda _{\beta }\left(
t\right) \phi _{\beta }\left( x,t\right)
\end{equation}
so the Heisenberg field operator becomes

\begin{equation}
\Psi \left( x,t\right) =a_{0}\left( t\right) \left[ \phi _{0}\left(
x,t\right) +\frac{1}{\sqrt{N}}\Lambda \left( x,t\right) \right]
\label{pcnop}
\end{equation}
we have the identities

\begin{equation}
\left( \phi _{0},\Lambda \right) =0
\end{equation}

\begin{equation}
\left\langle a_{0}^{\dagger }a_{0}\Lambda \left( x,t\right) \right\rangle =0
\label{identity}
\end{equation}
The condensate occupation number is

\begin{equation}
N_{0}=N\left[ 1-\frac{\nu }{N}+o\left( N^{-1}\right) \right]  \label{noncond}
\end{equation}

\begin{equation}
\nu =\sum_{\beta \neq 0}\lambda _{\beta }^{\dagger }\lambda _{\beta }
\label{nu}
\end{equation}

We seek a solution of the Heisenberg equations of motion Eq. (\ref{Hei2}) of
the form

\begin{mathletters}
\begin{equation}
a_{0}=\sqrt{N}e^{-i\Theta \left( t\right) }\left[ \gamma +\frac{\gamma _{1}}{%
N^{1/2}}+\frac{\Delta \gamma }{N}\right]  \label{ansatz1}
\end{equation}

\end{mathletters}
\begin{equation}
\phi _{0}=\Phi +\frac{\Delta \Phi }{N}  \label{ansatz2}
\end{equation}

\begin{equation}
\Lambda =\Lambda _{0}+\frac{\Delta \Lambda }{N^{1/2}}  \label{ansatz3}
\end{equation}
where $\gamma $ is a constant $q$-number and $\gamma ^{\dagger }\gamma
=\gamma \gamma ^{\dagger }=1$. Expanding the canonical commutation relations
we get

\begin{equation}
\left[ \gamma ,\gamma _{1}^{\dagger }\right] +\left[ \gamma _{1},\gamma
^{\dagger }\right] =\left[ \Lambda _{0},\gamma ^{\dagger }\right] =0
\label{etccr1}
\end{equation}

\begin{equation}
\left[ \gamma _{1},\gamma _{1}^{\dagger }\right] +\left[ \gamma ,\Delta
\gamma ^{\dagger }\right] +\left[ \Delta \gamma ,\gamma ^{\dagger }\right] =1
\label{etccr2}
\end{equation}

The expansion of the Heisenberg equations in inverse powers of $N$ yields at
leading order the Gross - Pitaievsky equation (GPE) for $\Phi $

\begin{equation}
0=i\frac{\partial \Phi }{\partial t}-H\Phi -u\Phi ^{*}\Phi ^{2}+\mu \Phi
\label{GPE}
\end{equation}
with a possibly time-dependent chemical potential

\begin{equation}
\mu =\frac{d\Theta }{dt}  \label{chempot}
\end{equation}
The actual value of $\mu $ is derived by consistency with the normalization
of $\Phi $. The next-to-leading order (NLO) terms read

\begin{equation}
0=i\gamma ^{\dagger }\frac{d\gamma _{1}}{dt}\Phi +i\frac{\partial \Lambda
_{0}}{\partial t}-H\Lambda _{0}-u\left[ \left( 2\Phi ^{*}\Phi -\frac{\mu }{u}%
\right) \Lambda _{0}+\Phi ^{2}\Lambda _{0}^{\dagger }\right]
\end{equation}
Projecting over $\Phi $ we get

\begin{equation}
0=i\gamma ^{\dagger }\frac{d\gamma _{1}}{dt}-\left( \Phi ,u\left[ \Phi
^{*}\Phi \Lambda _{0}+\Phi ^{2}\Lambda _{0}^{\dagger }\right] \right)
\label{gamma1}
\end{equation}
and finally

\begin{equation}
0=i\frac{\partial \Lambda _{0}}{\partial t}-H\Lambda _{0}-u\left( \Phi
^{*}\Phi \right) \Lambda _{0}+\mu \Lambda _{0}-uQ\left[ \Phi ^{*}\Phi
\Lambda _{0}+\Phi ^{2}\Lambda _{0}^{\dagger }\right]  \label{transverse}
\end{equation}
where

\begin{equation}
Q\left[ f\right] =f-\Phi \left( \Phi ,f\right)  \label{projop}
\end{equation}

Observe that the equation for $\gamma _{1}$ is consistent with the
requirement that $N_{0}/N=1-O\left( N^{-1}\right) ,$ since

\begin{equation}
\frac{d}{dt}\left[ \gamma ^{\dagger }\gamma _{1}+\gamma _{1}^{\dagger
}\gamma \right] =0  \label{consistency}
\end{equation}
so we may impose the condition $\gamma ^{\dagger }\gamma _{1}+\gamma
_{1}^{\dagger }\gamma =0.$

We see that these equations do not determine $\gamma .$ Since the one body
density matrix is independent of it, $\gamma $ may be chosen freely. The
simplest choice is $\gamma =1,$ in which case Eq. (\ref{etccr2}) reduces to

\begin{equation}
\left[ \gamma _{1},\gamma _{1}^{\dagger }\right] =1\qquad \left( \gamma
=1\right)  \label{etccr3}
\end{equation}

With this choice, the requirement that $\gamma _{1}+\gamma _{1}^{\dagger }=0 
$ must be understood as a restriction on the allowed states, rather than as
a strong identity, since it would conflict with Eq. (\ref{etccr3}). We shall
discuss this issue in more detail in the next section.

\section{Gauge invariant effective action for Bose-Einstein condensates}

Consider the vacuum persistence amplitude for a theory with classical action 
$S$ as in Eq. (\ref{action}), expressed as a path integral over paths with
total particle number $N$

\begin{equation}
Z_{0}=\int D\Psi \;e^{iS/\hbar }\prod_{t}\delta \left[ \left( \int
d^{d}x\;\Psi ^{\dagger }\Psi \right) -N\right] .
\end{equation}
Exponentiate the delta functions

\begin{equation}
Z_{0}=\int D\Psi D\mu _{q}\;e^{iS_{\mu }/\hbar }
\end{equation}

\begin{equation}
S_{\mu }=S+\hbar \int d^{d+1}x\;\mu _{q}\left( t\right) \left[ \Psi
^{\dagger }\Psi -\frac{N}{V}\right]
\end{equation}
Observe that now the path integral is redundant, since we may transform the
fields as in Eq. (\ref{local}). We may fix the redundancy by factoring out
the gauge group. Choose some function $f_{\theta }=f\left[ \mu _{q\theta
},\Psi _{\theta },\Psi _{\theta }^{\dagger }\right] ,$ such that $df_{\theta
}/d\theta \neq 0.$ Then

\begin{equation}
1=\int \frac{df_{\theta }}{d\theta }d\theta \;\delta \left( f_{\theta
}-c\right)
\end{equation}
Inserting this into the vacuum persistence amplitude and average over $c$
with a weight $e^{ic^{2}/2\sigma }$ we get

\begin{equation}
Z_{0}=\Theta \int D\Psi D\mu _{q}\;e^{iS_{\mu _{q},\sigma }/\hbar }Det\left[ 
\frac{\delta f_{\theta }}{\delta \theta }\right] _{\theta =0}
\end{equation}
where

\begin{equation}
\Theta =\int D\theta
\end{equation}
is the volume of the gauge group we wish to factor out.

\begin{equation}
S_{\mu ,\sigma }=S+\hbar \int d^{d+1}x\;\mu _{q}\left( t\right) \left[ \Psi
^{\dagger }\Psi -\frac{N}{V}\right] +\frac{\hbar }{2\sigma }\int
dt\;f_{0}^{2}  \label{gaugefixed}
\end{equation}
The determinant is expressed as a path integral over Grassmann fields $\zeta
,\eta $

\begin{equation}
Det\left[ \frac{\delta f_{\theta }}{\delta \theta }\right] _{\theta =0}=\int
D\zeta D\eta \;e^{-\frac{1}{\hbar }\int dt\;\zeta \frac{\delta f_{\theta }}{%
\delta \theta }\eta }
\end{equation}

Thus, we have a theory of fields $\Psi ,$ $\Psi ^{\dagger },$ $\mu _{q}$, $%
\zeta $ and $\eta .$ Suppose we wish now to compute the one loop effective
action for this theory. We redefine $\Psi =\bar{\Phi}+\Psi _{q}$ as in Eq. (%
\ref{mfs}). Similarly redefine $\Psi ^{\dagger }=\bar{\Phi}^{*}+\Psi
_{q}^{\dagger }$, and $\mu _{q}=\bar{\mu}_{q}+\lambda $. Doing so introduces
an undesirable cross term

\begin{equation}
\hbar \int d^{d+1}x\;\lambda \left( t\right) \left[ \bar{\Phi}^{*}\Psi _{q}+%
\bar{\Phi}\Psi _{q}^{\dagger }\right]
\end{equation}
To eliminate this term, we choose

\begin{equation}
f=-\sigma \left( \mu _{q}-\bar{\mu}_{q}\right) +\int d^{d}x\;\left[ \bar{\Phi%
}^{*}\Psi +\bar{\Phi}\Psi ^{\dagger }-2\left| \bar{\Phi}\right| ^{2}\right]
\label{gft}
\end{equation}
The ghost action now becomes

\begin{equation}
S_{ghost}=i\int dt\;\zeta \left[ -\sigma \frac{d}{dt}+\int d^{d}x\;\left( 
\bar{\Phi}^{*}\Psi _{q}-\bar{\Phi}\Psi _{q}^{\dagger }\right) \right] \eta
\label{gt}
\end{equation}
and it decouples at one-loop order.

\subsection{Inverse particle number as a small parameter}

The loop expansion is a formal development in which Feynman graphs are
classified according to their topology. This has strong heuristic value but
is hard to estimate a priori which order in perturbation theory must be
reached to guarantee any prescribed accuracy. The inverse of the total
particle number $N$ provides a small parameter with a clear physical
meaning, and so it is convenient to organize the perturbative expansion as a
development in inverse powers of $N.$ As we shall see, this may be achieved
through a Stratonovich transformation of the interaction term.

Let us begin by rewriting the classical action as (from now on, we assume $%
\hbar =1$)

\begin{equation}
S_{N}=\int dtd^{d}x\;\left\{ i\Psi ^{\dagger }D_{t}\Psi -\Psi ^{\dagger
}H\Psi -\frac{u}{2N}\Psi ^{\dagger 2}\Psi ^{2}-N\frac{\mu _{q}}{V}\right\}
\end{equation}
with $\mu _{q}$ as connection in the covariant derivative

\begin{equation}
D_{t}=\frac{\partial }{\partial t}-i\mu _{q}
\end{equation}
We have assumed the nonlinear term scales with $N$ in a particular way ($%
U\rightarrow u/N$) to make sure the large $N$ limit is well defined.

We add a term

\begin{equation}
\frac{u}{2N}\int dtd^{d}x\;\left( \Psi ^{\dagger 2}-\frac{N}{u}\Omega
^{\dagger }\right) \left( \Psi ^{2}-\frac{N}{u}\Omega \right)
\end{equation}
to the action, whose only effect is to multiply the generating functional by
a constant factor, and integrate over the new auxiliary fields $\Omega $ and 
$\Omega ^{\dagger }$. The action now reads 
\begin{equation}
S_{N}=\int dtd^{d}x\;\left\{ i\Psi ^{\dagger }D_{t}\Psi -\Psi ^{\dagger
}H\Psi -\frac{1}{2}\left( \Omega ^{\dagger }\Psi ^{2}+\Psi ^{\dagger
2}\Omega \right) +\frac{N}{2u}\Omega ^{\dagger }\Omega -N\frac{\mu _{q}}{V}%
\right\}
\end{equation}
We add to this action the gauge fixing term as in Eq. (\ref{gaugefixed}).
The gauge fixing condition is given by Eq. (\ref{gft}), where the gauge
parameter is chosen to be $\sigma =Ns,$ and the ghost term Eq. (\ref{gt}).

Finally we rescale $\bar{\Phi}=\sqrt{N}\phi $, $\bar{\Phi}^{*}=\sqrt{N}\phi
^{*}$ and $\Psi _{q}=\sqrt{N}{\tilde{\psi}}$, $\Psi _{q}^{\dagger }=\sqrt{N}{%
\tilde{\psi}}^{\dagger }$, whereby we get

\begin{equation}
\Psi =\sqrt{N}\left[ \phi +\tilde{\psi}\right]  \label{smfs}
\end{equation}

\begin{equation}
S_{N}\left[ \Psi ,\Omega ,\mu _{q},\zeta ,\eta ,\bar{\Phi}\right]
=NS_{1}\left[ \phi +{\tilde{\psi}},\Omega ,\mu _{q},\zeta ,\eta ,\phi \right]
\end{equation}
Henceforth, we shall write $S_{1}\equiv S$.

In the rescaled theory there are only cubic interactions; all vertices scale
like $N$, and all propagators scale as $N^{-1}$. It follows that the power
of $N$ for any given graph is simply $1-\ell $, where $\ell $ is the number
of loops.

\subsection{Leading order equations of motion}

Introduce background fields $\bar{\Omega}$ and $\bar{\mu}_{q}$ for the
fields $\Omega =\bar{\Omega}+\omega $, $\mu _{q}=\bar{\mu}_{q}+\lambda $. To
leading order in $N$, the theory reduces to the classical one, with
equations of motion

\begin{equation}
i\bar{D}_{t}\phi -H\phi -\bar{\Omega}\phi ^{*}=0  \label{mfe}
\end{equation}

\begin{equation}
\int d^{d}x\;\phi ^{\ast }\phi =1
\end{equation}

\begin{equation}
\phi ^{2}-\frac{\bar{\Omega}}{u}=0
\end{equation}
where $\bar{D}\equiv \frac{\partial }{\partial t}-i\bar{\mu}_{q}$ is the
covariant derivative evaluated with $\bar{\mu}_{q}$ as connection. To make
contact with the PNC formalism above, we assume (cfr. Eqs. (\ref{ansatz1})
and (\ref{ansatz2}))

\begin{equation}
\phi =e^{-i\Theta \left( t\right) }\left[ \Phi +\frac{\Delta \phi }{N}\right]
\label{ansatz4}
\end{equation}
and $\bar{\mu}_{q}=0.$ We therefore get the GPE Eq. (\ref{GPE}) for $\Phi ,$
with $\mu =d\Theta /dt$ as in Eq. (\ref{chempot}) above.

\subsection{The fluctuation fields at leading order}

To compute the first order corrections, namely, terms of order $N^{0}$ in
the effective action, we must keep quadratic terms in the fluctuations,
namely

\begin{equation}
S_{N}\left[ \Psi ,\Omega ,\mu _{q},\zeta ,\eta ,\bar{\Phi}\right] =NS\left[
\phi ,\bar{\Omega},\bar{\mu}_{q},\zeta ,\eta ,\phi \right] +NS_{quad}
\end{equation}

\begin{eqnarray}
S_{quad} &=&\int dtd^{d}x\;\left\{ i{\tilde{\psi}}^{\dagger }\bar{D}_{t}{%
\tilde{\psi}}-{\tilde{\psi}}^{\dagger }H{\tilde{\psi}}-\frac{1}{2}\left( 
\bar{\Omega}^{\dagger }{\tilde{\psi}}^{2}+{\tilde{\psi}}^{\dagger 2}\bar{%
\Omega}\right) \right.  \nonumber \\
&&\left. -\left[ \phi \omega ^{\dagger }{\tilde{\psi}}+\phi ^{*}{\tilde{\psi}%
}^{\dagger }\omega \right] +\frac{1}{2u}\omega ^{\dagger }\omega \right\} 
\nonumber \\
&&+\frac{1}{2s}\int dt\;\left\{ \int d^{d}x\;\left[ \phi ^{*}{\tilde{\psi}}+{%
\tilde{\psi}}^{\dagger }\phi \right] \right\} ^{2}+s\int dt\;\left\{ \frac{%
\lambda ^{2}}{2}-i\zeta \frac{d\eta }{dt}\right\}
\end{eqnarray}

The ghost and gauge action decouple to lowest order, but they must be
included at higher orders. To eliminate the $\omega ${$\tilde{\psi}$} cross
terms, we redefine

\begin{equation}
\omega =\varpi +2u\phi {\tilde{\psi}}
\end{equation}
to get

\begin{eqnarray}
S_{quad} &=&\int dtd^{d}x\;\left\{ i{\tilde{\psi}}^{\dagger }\bar{D}_{t}%
\tilde{\psi}-{\tilde{\psi}}^{\dagger }H{\tilde{\psi}}-\frac{1}{2}\left( \bar{%
\Omega}^{\dagger }{\tilde{\psi}}^{2}+{\tilde{\psi}}^{\dagger 2}\bar{\Omega}%
\right) -2u\phi ^{*}\phi {\tilde{\psi}}^{\dagger }{\tilde{\psi}}+\frac{1}{2u}%
\varpi ^{\dagger }\varpi \right\}  \nonumber \\
&&\ +\frac{1}{2s}\int dt\;\left\{ \int d^{d}x\;\left[ \phi ^{*}{\tilde{\psi}}%
+{\tilde{\psi}}^{\dagger }\phi \right] \right\} ^{2}+s\int dt\;\left\{ \frac{%
\lambda ^{2}}{2}-i\zeta \frac{d\eta }{dt}\right\}
\end{eqnarray}

The $\varpi $, $\lambda $ and the ghost fields are decoupled and play no
further role at NLO. Making one last scaling and a phase shift 
\begin{equation}
\tilde{\psi}=N^{-1/2}e^{-i\Theta }\psi ,
\end{equation}
we find that to leading order the field $\psi $ is a quantized linear field
obeying the Heisenberg equation of motion

\begin{equation}
iD_{t}\psi -H\psi -u\Phi ^{2}\psi ^{\dagger }-2u\Phi ^{*}\Phi \psi =\frac{%
-\Phi }{s}\int d^{d}y\;\left[ \Phi ^{*}\psi +\psi ^{\dagger }\Phi \right]
\left( y,t\right)  \label{hei3}
\end{equation}
and commutation relations

\begin{equation}
\left[ \psi \left( x,t\right) ,\psi ^{\dagger }\left( y,t\right) \right]
=\delta \left( x-y\right)
\end{equation}
We have assumed $\bar{\mu}_{q}=0;$ $D_{t}\equiv \frac{\partial }{\partial t}%
-i\mu $ is the covariant derivative with $\mu $ as connection.

Let us split the fluctuation field into its components along the mean field
(longitudinal) and orthogonal to it (transverse)

\begin{equation}
\psi \left( x,t\right) =\psi _{0}\left( t\right) \Phi \left( x,t\right)
+\psi _{\perp }\left( x,t\right)  \label{split}
\end{equation}
where $\psi _{0}=\left( \Phi ,\psi \right) $ and $\psi _{\perp }=Q\psi $
(cfr. Eq. (\ref{projop})). The ETCCR imply

\begin{equation}
\left[ \psi _{0},\psi _{0}^{\dagger }\right] =1
\end{equation}

\begin{equation}
\left[ \psi _{0},\psi _{\perp }^{\dagger }\right] =0
\end{equation}

\begin{equation}
\left[ \psi _{\perp }\left( x,t\right) ,\psi _{\perp }^{\dagger }\left(
y,t\right) \right] =Q\delta \left( x-y\right)
\end{equation}

We substitute Eq. (\ref{split}) into the wave equation for the fluctuations
Eq. (\ref{hei3}); using also the GPE Eq. (\ref{GPE})

\begin{equation}
0=i\frac{d\psi _{0}}{dt}\Phi +A\left[ \psi _{0}+\psi _{0}^{\dagger }\right]
-C\left[ \psi _{\perp }\right]  \label{hei4}
\end{equation}
where

\begin{equation}
A=\frac{\Phi }{s}-u\Phi ^{*}\Phi ^{2}
\end{equation}

\begin{equation}
C\left[ \psi _{\perp }\right] =-iD_{t}\psi _{\perp }+H\psi _{\perp }+u\Phi
^{2}\psi _{\perp }^{\dagger }+2u\Phi ^{*}\Phi \psi _{\perp }  \label{c}
\end{equation}

Projecting over $\Phi $ we get the equation for the longitudinal mode

\begin{equation}
i\frac{d\psi _{0}}{dt}+A_{0}\left[ \psi _{0}+\psi _{0}^{\dagger }\right]
=C_{0}\left[ \psi _{\perp }\right]  \label{long}
\end{equation}
where

\begin{equation}
A_{0}=\left( \Phi ,A\right)
\end{equation}

\begin{equation}
C_{0}\left[ \psi _{\perp }\right] =\left( \Phi ,C\left[ \psi _{\perp
}\right] \right) =u\left( \Phi ,\left[ \Phi ^{2}\psi _{\perp }^{\dagger
}+\Phi ^{*}\Phi \psi _{\perp }\right] \right)  \label{c0}
\end{equation}

Now observe that both $A_{0}$ and $C_{0}\left[ \psi _{\perp }\right] $ are
real, so writing

\begin{equation}
\psi _{0}=\xi +i\eta
\end{equation}
with ETCCR

\begin{equation}
\left[ \xi ,\eta \right] =\frac{i}{2}  \label{etccr0}
\end{equation}
we get

\begin{equation}
\frac{d\xi }{dt}=0  \label{realpart}
\end{equation}

\begin{equation}
\frac{d\eta }{dt}=2A_{0}\xi -C_{0}\left[ \psi _{\perp }\right]
\label{impart}
\end{equation}

We realize that Eq. (\ref{realpart}) is the counterpart to Eq. (\ref
{consistency}), so to match the GIEA and PNC approaches we set $\xi =0.$
Comparing Eq. (\ref{impart}) to Eq. (\ref{gamma1}), we see that the
longitudinal fluctuation field is simply $\psi _{0}=\gamma _{1},$ where the
choice $\gamma =1$ has been made. As discussed in the last section, we do
not regard $\xi =0$ as a strong condition, but rather as a restriction on
allowed physical states. We shall discuss this issue in the next subsection.

The equation for the orthogonal component $\psi _{\perp }$ is now 
\begin{equation}
QC\left[ \psi _{\perp }\right] =C\left[ \psi _{\perp }\right] -\Phi
C_{0}\left[ \psi _{\perp }\right] =0  \label{psitranseq}
\end{equation}
where $C$ and $C_{0}$ are defined in Eqs. (\ref{c}) and (\ref{c0}),
respectively. Since this is equivalent to Eq. (\ref{transverse}), we
identify $\psi _{\perp }=\Lambda _{0}.$

\subsection{Gauge fixing dependence of the two point functions}

To complete the evaluation of the next-to-leading order (NLO) corrections to
the mean fields and the one-body density matrix, we need the LO two point
functions of the theory, that is, the expectation values of the products of
two fluctuation fields. In this subsection we shall discuss the important
issue of whether these two-point functions are gauge-fixing dependent, that
is, whether they depend on the parameter $s$. Of course, physical
observables, the one - body density matrix among them, must be gauge fixing
independent.

Observe that $s$ has disappeared from the Heisenberg equation for $\psi
_{\perp }$ Eq. (\ref{psitranseq}), so the two point functions built from it
are automatically gauge fixing independent. As for the two point functions
involving the longitudinal modes $\xi $ and $\eta ,$ we translate the
requirement $\xi =0$ into

\begin{equation}
\left\langle \xi ^{2}\right\rangle =\left\langle \xi \psi _{\perp
}\right\rangle =\left\langle \xi \eta +\eta \xi \right\rangle =0
\label{vanishing}
\end{equation}
while the ETCCR means

\begin{equation}
\left\langle \xi \eta -\eta \xi \right\rangle =\frac{i}{2}
\label{nonvanishing}
\end{equation}
Eq. (\ref{vanishing}) implies

\begin{equation}
\frac{d}{dt}\left\langle \eta \left( t\right) \psi _{\perp }\left(
y,t^{\prime }\right) \right\rangle =-C_{0}\left[ \left\langle \psi _{\perp
}\left( \;,t\right) \psi _{\perp }\left( y,t^{\prime }\right) \right\rangle
\right]
\end{equation}
We see that there are nontrivial correlations between the longitudinal and
the transverse quantum field, but they are gauge fixing independent.

Finally

\begin{equation}
\frac{d}{dt}\left\langle \eta \left( t\right) \eta \left( t^{\prime }\right)
\right\rangle =\frac{i}{2}A_{0}\left( t\right) -C_{0}\left[ \left\langle
\psi _{\perp }\left( \;,t\right) \eta \left( t^{\prime }\right)
\right\rangle \right] 
\end{equation}

\begin{equation}
\frac{d}{dt^{\prime }}\left\langle \eta \left( t\right) \eta \left(
t^{\prime }\right) \right\rangle =-\frac{i}{2}A_{0}\left( t^{\prime }\right)
-C_{0}\left[ \left\langle \eta \left( t\right) \psi _{\perp }\left(
\;,t^{\prime }\right) \right\rangle \right] 
\end{equation}
These two equations show that we may write

\begin{equation}
\left\langle \eta \left( t\right) \eta \left( t^{\prime }\right)
\right\rangle =\left\langle \eta \left( t\right) \eta \left( t^{\prime
}\right) \right\rangle _{0}+\frac{i}{2}\int_{t^{\prime }}^{t}d\tau
\;A_{0}\left( \tau \right)
\end{equation}
where $\left\langle \eta \left( t\right) \eta \left( t^{\prime }\right)
\right\rangle _{0}$ is gauge fixing independent. In particular, the gauge
fixing dependence disappears in the coincidence limit $t^{\prime }=t$.

We may now proceed to evaluate the NLO mean fields.

\subsection{The mean fields at next to leading order}

We may write the GIEA to one loop order as

\begin{equation}
\Gamma _{GI}\left[ \phi ,\bar{\mu},\bar{\Omega}\right] =NS\left[ \phi ,\bar{%
\Omega},\bar{\mu}_{q},0,0,\phi \right] +\Gamma _{1}\left[ \phi ,\bar{\mu}%
_{q},\bar{\Omega}\right] 
\end{equation}

\begin{equation}
\Gamma _{1}=-i\ln \left[ \int D\psi D\psi ^{\dagger }\;e^{iS_{quad}\left[
\psi ,\psi ^{\dagger }\right] }\right]  \label{intrep}
\end{equation}

To NLO we write the mean fields as

\begin{equation}
\phi =e^{-i\Theta \left( t\right) }\left[ \Phi +\frac{\phi ^{\left( 1\right)
}}{N}+o\left( N^{-1}\right) \right]
\end{equation}

\begin{equation}
\bar{\Omega}=e^{-2i\Theta \left( t\right) }\left[ u\Phi ^{2}+\frac{\bar{%
\Omega}^{\left( 1\right) }}{N}+o\left( N^{-1}\right) \right]
\end{equation}

\begin{equation}
\bar{\mu}_{q}=\frac{\mu ^{\left( 1\right) }}{N}
\end{equation}

We also split $\phi ^{\left( 1\right) }$ into components along the LO mean
field (longitudinal) and perpendicular to it (transverse)

\begin{equation}
\phi ^{\left( 1\right) }=\phi _{0}^{\left( 1\right) }\Phi +\phi _{\perp
}^{\left( 1\right) }
\end{equation}

The equations of motion for the NLO terms become

\begin{eqnarray}
0 &=&iD_{t}\phi ^{\left( 1\right) }-H\phi ^{\left( 1\right) }-u\Phi ^{2}\phi
^{\left( 1\right) *}-\bar{\Omega}^{\left( 1\right) }\Phi ^{*}-2u\left\langle
\psi ^{\dagger }\psi \right\rangle \Phi  \nonumber \\
&&+\frac{1}{s}\left\langle \left\{ \psi _{0}\left( t\right) \Phi \left(
x,t\right) +\psi _{\perp }\left( x,t\right) ,\xi \right\} \right\rangle +\mu
^{\left( 1\right) }\Phi  \label{nlo1}
\end{eqnarray}

\begin{equation}
\phi _{0}^{\left( 1\right) }+\phi _{0}^{\left( 1\right) *}+\left\langle \psi
_{0}^{\dagger }\psi _{0}\right\rangle +\nu =0  \label{nlo2}
\end{equation}

\begin{equation}
\Phi \left( 2\phi ^{\left( 1\right) }+\left\langle \left\{ \psi _{0},\psi
_{\perp }\right\} \right\rangle \right) +\left\langle \psi
_{0}^{2}\right\rangle \Phi ^{2}+\left\langle \psi _{\perp }^{2}\right\rangle
-\frac{\bar{\Omega}^{\left( 1\right) }}{u}=0  \label{nlo3}
\end{equation}
where we identified $\psi _{\perp }=\Lambda _{0}$ and introduced $\nu $ from
Eq. (\ref{nu}).

In view of the analysis in the previous subsection, the explicitly $s$%
-dependent term in Eq. (\ref{nlo1}) vanishes. All other two-point functions
are in the coincidence limit, and so they are gauge fixing independent. We
may conclude that the NLO correction to the mean fields themselves is gauge
fixing independent. This is the only property of these corrections we shall
need in the remainder.

\subsection{The one-body density matrix at next to leading order}

From the analysis above, we see that in the GIEA approach we obtain the
Heisenberg field operator for the bosonic field as

\begin{equation}
\Psi =\sqrt{N}e^{-i\Theta \left( t\right) }\left\{ \left[ 1+\frac{\psi _{0}}{%
\sqrt{N}}+\frac{\phi _{0}^{\left( 1\right) }}{N}\right] \Phi +\frac{\psi
_{\perp }}{\sqrt{N}}+\frac{\phi _{\perp }^{\left( 1\right) }}{N}\right\}
\end{equation}
Observe that this is also the result of the canonical PNC approach Eq. (\ref
{pcnop}) under the choice $\gamma =1.$ Since it is observable, $\sigma _{m}$
must be explicitly gauge independent.

We get, using the NLO equations of motion Eq. (\ref{nlo2}),

\begin{equation}
\sigma _{m}\left( x,y\right) =\sigma _{0}\left( x,y\right) +\delta \sigma
_{m}\left( x,y\right)
\end{equation}

\begin{equation}
\sigma _{0}\left( x,y\right) =N\Phi ^{*}\left( x\right) \Phi \left( y\right)
\left[ 1-\frac{\nu }{N}\right]
\end{equation}

\begin{eqnarray}
\delta \sigma _{m}\left( x,y\right) &=&\Phi ^{*}\left( x\right) \left( \phi
_{\perp }^{\left( 1\right) }\left( y\right) +\left\langle \psi _{0}^{\dagger
}\psi _{\perp }\left( y\right) \right\rangle \right)  \nonumber \\
&&+\left( \phi _{\perp }^{\left( 1\right) *}\left( x\right) +\left\langle
\psi _{\perp }^{\dagger }\left( x\right) \psi _{0}\right\rangle \right) \Phi
\left( y\right) +\left\langle \psi _{\perp }^{\dagger }\left( x\right) \psi
_{\perp }\left( y\right) \right\rangle
\end{eqnarray}

To diagonalize this, we simply think of $\delta \sigma _{m}$ as a
perturbation of $\sigma _{0}$. $\sigma _{0}$ clearly admits $\Phi $ as an
eigenvector with eigenvalue $N-\nu $. Since $\left( \Phi ,\delta \sigma
_{m}\Phi \right) =0,$ there is no correction to this eigenvalue at first
order. This means that the condensate occupation number is $N_{0}=N\left[
1-\nu /N+o\left( N^{-1}\right) \right] $, as in the canonical approach, Eq. (%
\ref{noncond}). The actual condensate wave function is

\begin{equation}
\phi _{0}=\Phi +\frac{1}{N}\left( \phi _{\perp }^{\left( 1\right) }\left(
y\right) +\left\langle \psi _{0}^{\dagger }\psi _{\perp }\left( y\right)
\right\rangle \right)
\end{equation}
which is not the same as the NLO mean field. Both the condensate occupation
number and wave function are explicitly gauge independent, as expected.

Finally we mention that, in spite of there being nontrivial mean fields, the
expectation value of the bosonic field operators is zero, as it must be in a
finite system. This comes about because the field operator transforms a
physical state into one that does not satisfy the particle number
constraint; for further discussion, see \cite{neqmott}.

\section{Discussions}

We have presented a functional formulation for cold atomic gases. This
condition enters as a constraint in the path integral, thus changing the
global $U\left( 1\right) $ symmetry of the model to a local (in time) one.
Therefore the theory must be quantized by the Fadeev - Popov method. We
derived a gauge-invariant effective action for the description of its
dynamics.

To make contact with the more familiar approaches, we draw attention to the
fact that the mean field is not necessarily identical with the condensate
wave function. However, once the mean field is found, it is easy to obtain
the condensate wave function as an expansion in inverse powers of particle
number.

The approach presented in this paper unveils the physical equivalence of
several proposals in the literature, which now can be seen as different
gauge (and gauge fixing condition) choices within the same theory. Thus, if
one takes the $s\rightarrow \infty $ limit, one eliminates the fluctuations
in the self-consistent, time-dependent chemical potential $\mu $ and the
ghost field, but places no restriction on quantum fluctuations of the atomic
field in the condensate mode. This is the road taken in \cite{MOR98} and 
\cite{ByHFB}. In the opposite $s\rightarrow 0$ limit one forces quantum
fluctuations in the condensate mode to vanish, but must include chemical
potential and ghost fluctuations in higher order perturbation theory. As we
can see, neither of these alternatives is the best in every conceivable
situation, which testifies to the value of a general formalism as presented
here. The other gain in this approach is of course to make contact with the
extensive body of work on gauge theories.

The more general nature of this new (gauge invariant) approach is the
appearance of a new parameter $s$. While explicitly present in the
intermediate steps it does not appear in the final result, as physical
predictions should not depend on it (gauge-fixing independence). As there is
considerable expertise in identifying the physically relevant, $s$ -
independent predictions of the theory, this is a lesser evil than its
appearance suggests \cite{KKR91}. After all this is as old a topic as
electromagnetism, told in the modern gauge theory language. \newline

As a first concrete application, in \cite{neqmott} the functional PNC
approach is used to study the one-body density matrix for a cold bosonic gas
in an optical lattice, near the superfluid - insulator transition. In this
application yet another gauge choice is made (in gauge theory parliance, a
simpler ``covariant'' gauge as oppossed to the ``background field'' gauges
used here). This reflects on the flexibility of the method to adapt to the
challenges of a concrete environment, knowing that physical equivalence is
guaranteed throughout by general theorems.\newline

\noindent\textbf{Acknowledgments} We thank Ana Maria Rey for helpful
discussions. EC acknowledges support from Universidad de Buenos Aires,
CONICET and ANPCyT (Argentina) and BLH from NSF grant PHY-0426696.

\newpage \appendix

\section{Brief survey of PNC proposals}

In this Appendix, we shall compare in some detail the PNC formalism
presented in Section II with those of \cite{GA59,GAR97,CD98}. We proceed in
chronological order of publication.

\subsection{The Girardeau - Arnowitt theory}

The Girardeau - Arnowitt (GA) theory \cite{GA59} stands apart from the other
two proposals to be discussed because its goal is to characterize the BEC
ground state within a variational approximation, rather than a dynamical
theory. Only the homogeneous case is discussed, so linear momentum is a good
quantum number for one-particle states. Finally, GA do not perform a large $N
$ expansion; they aim to describe the BEC regime even when depletion is
large.

The basic insight of the GA theory is that in the Bogoliubov theory the
ground state is seen as a coherent superposition of states with different
numbers of particle pairs (a particle of momentum $p$ correlated to one of
momentum $-p$) and thus different total particle numbers (see any textbook
presentation, e. g. \cite{HUA87}). GA describe the ground state as a
coherent superposition of particle - hole pairs (a particle of momentum $p$,
another one of momentum $-p,$ and two particles less in the condensate), all
of them having the same total particle number. A particle - hole is
destroyed by the operator $c_{p}=a_{p}\beta _{0}^{\dagger }$ \cite{G98},
where $a_{p}$ destroys a particle of momentum $p$ and

\begin{equation}
\beta _{0}=\left( N_{0}+1\right) ^{-1/2}a_{0}
\end{equation}
where we follow the updated formulation of \cite{G98}. In terms of the
notation of Section II, Eqs. (\ref{ansatz1}), (\ref{ansatz2}) and (\ref
{ansatz3}) we have (we assume $\gamma =1$)

\begin{mathletters}
\begin{eqnarray}
\beta _{0} &=&\sqrt{\frac{N}{N_{0}+1}}e^{-i\Theta \left( t\right) }\left[ 1+%
\frac{\gamma _{1}}{N^{1/2}}+\frac{\Delta \gamma }{N}\right]   \nonumber \\
&\sim &e^{-i\Theta \left( t\right) }\left[ 1+\frac{\gamma _{1}}{N^{1/2}}+%
\frac{\Delta \gamma ^{\prime }}{N}\right] 
\end{eqnarray}
From Eq. (\ref{newop}), the particle - hole destruction operator is

\end{mathletters}
\begin{eqnarray}
c_{p} &=&\frac{1}{\sqrt{N}}\lambda _{p}N_{0}\left( N_{0}+1\right) ^{-1/2} 
\nonumber \\
&=&\lambda _{p}\frac{N_{0}}{\sqrt{N\left( N_{0}+1\right) }}\sim \lambda
_{p}+O\left( N^{-1}\right)
\end{eqnarray}

In the homogeneous case $\Phi =\phi _{0}=1/\sqrt{V}.$ Therefore Eq. (\ref
{gamma1}) gives $\gamma _{1}=$ constant, the Gross - Pitaievsky equation (%
\ref{GPE}) yields $\mu =u/V$ and Eq. (\ref{transverse}) becomes

\begin{equation}
0=i\frac{\partial \lambda _{p}}{\partial t}-\frac{p^{2}}{2m}\lambda _{p}-%
\frac{u}{V}\left[ \lambda _{p}+\lambda _{p}^{\dagger }\right]
\end{equation}
which also follows from the GA Hamiltonian at leading order (see Eq. (14) in 
\cite{G98})

These formulae provide the translation between the GA theory and our
formalism.

\subsection{The Gardiner theory}

Gardiner's theory \cite{GAR97} is also a systematic large $N$ expansion of
the Heisenberg equations of motion, after introducing an ansatz equivalent
to Eq. (\ref{pcnop}) for the field operator. The condensate wave function is
not introduced as an eigenfunction of the one-body density matrix, but
rather kept as an unknown. The Gross - Pitaievsky equation Eq. (\ref{GPE})
is derived as a consistency condition, because if it does not hold, then the
time derivative of the non-condensate field acquires a term of order $%
N^{1/2},$ thus invalidating the large $N$ expansion. This procedure is
sufficient for a NLO discussion as carried out in this paper, but the
condensate - non condensate separation becomes ambiguous at higher orders,
in which case the Castin and Dum proposal \cite{CD98} may be preferable.

\subsection{The Castin and Dum theory}

As we have stated in the Introduction, our presentation of the PNC method in
Section II follows the Castin and Dum theory \cite{CD98}. The only
difference is that Castin and Dum define the non-condensate operator as (in
the notation of Section II)

\begin{equation}
\Lambda _{ex}\left( x,t\right) =\frac{N_{0}}{N}\Lambda \left( x,t\right)
\end{equation}
This makes Eq. (\ref{identity}) simpler

\begin{equation}
\left\langle \Lambda _{ex}\left( x,t\right) \right\rangle =0
\end{equation}
allowing for a clearcut separation between the condensate and noncondensate
contributions to the equations of motion, but the expression of the
Heisenberg field in terms of $\Lambda _{ex}\left( x,t\right) $ is more
involved than in terms of $\Lambda \left( x,t\right) .$ Otherwise, both
formulations are fully equivalent.

For further developments within the PNC formulation, see \cite{MOR03,DS03}.

\section{The gauge invariant effective action}

In this section, we shall review the theory of the GIEA. We begin by
introducing the effective action in a general non-gauge theory.

\subsection{The effective action \protect\cite{Jac74}}

Consider a theory of fields $\Psi ^i$ whose dynamics is described by an
action $S\left[ \Psi ^i\right].$ The evolution of the fields from some
initial to some final times (usually minus and plus infinity, respectively)
can be represented by the vacuum persistence amplitude defined as the
amplitude for the initial vacuum state to evolve into the OUT vacuum state:

\begin{equation}
Z\left[ 0\right] =\left\langle 0out\left| 0in\right. \right\rangle
=e^{iW\left[ 0\right] }=\int D\Psi ^i\;e^{iS\left[ \Psi ^i\right] }
\label{zetacero}
\end{equation}
where we set $\hbar =1$ and adopt the DeWitt convention that a single index $%
i$ denotes both space-time and internal indices, and repeated indices denote
integration over space-time and summation over internal indices. More
general initial and final states may be considered, but this shall be enough
for our purposes.

If we allow the fields to interact with external $c$-number sources $J_{i}$,
then the persistence amplitude becomes the generating functional. Its
Legendre transform is the effective action, which obeys

\begin{equation}
e^{i\Gamma \left[ \Phi \right] }=\int D\Psi ^i\;e^{i\left( S\left[ \Psi
^i\right] -\Gamma _{,i}\left( \Psi ^i-\Phi ^i\right) \right) }
\end{equation}
Observe that

\begin{equation}
e^{-i\Gamma \left[ \Phi \right] }\int D\Psi ^i\;\Psi ^ie^{i\left( S\left[
\Psi ^i\right] -\Gamma _{,i}\left( \Psi ^i-\Phi ^i\right) \right) }=\Phi ^i
\end{equation}
Consider fluctuations $\psi ^i$ around the background fields $\Phi ^i$ with $%
\Psi ^i=\Phi ^i+\psi ^i$, and expand the action for $\psi ^i$ to second
order in $\psi ^i$:

\begin{equation}
S\left[ \Phi +\psi \right] \sim S\left[ \Phi \right] +S_{,i}\psi ^i+\frac
12S_{,ij}\psi ^i\psi ^j
\end{equation}
Then

\begin{equation}
\Gamma \left[ \Phi \right] =S\left[ \Phi \right] +\delta \Gamma
\end{equation}
where 
\begin{equation}
e^{i\delta \Gamma \left[ \Phi \right] }=\int D\psi ^i\;e^{i\left( \frac
12S_{,ij}\psi ^i\psi ^j-\delta \Gamma _{,i}\psi ^i\right) }
\end{equation}
To leading order, $\delta \Gamma _{,i}$ on the right hand side may be
neglected, then the one-loop effective action is given by

\begin{equation}
\delta \Gamma =\frac i2\ln \;\mathrm{Det}\;\left[ S,_{ij}\right] ^{-1}
\end{equation}

\subsection{The effective action for gauge theories \protect\cite
{KuOj,DeWitt}}

Let us consider now the case where the classical action is invariant under
an infinitesimal gauge transformations parametrized by $\varepsilon ^\alpha $

\begin{equation}
\Psi ^i\rightarrow \Psi ^i\left[ \varepsilon \right] =\Psi ^i+\delta \Psi ^i
\end{equation}

\begin{equation}
\delta \Psi ^i=Q_\alpha ^i\left[ \Psi ^i\right] \varepsilon ^\alpha
\end{equation}

\begin{equation}
S_{,i}Q_\alpha ^i=0.
\end{equation}
We assume the gauge transformations are linear

\begin{equation}
Q_{\alpha ,jk}^{i}=0,
\end{equation}
that they form a group

\begin{equation}
Q_{\alpha ,j}^{i}Q_{\beta }^{j}-Q_{\beta ,j}^{i}Q_{\alpha }^{j}=C_{\alpha
\beta }^{\gamma }Q_{\gamma }^{i}
\end{equation}
and are volume-preserving

\begin{equation}
Q_{\alpha ,i}^i=0
\end{equation}
The structure constants satisfy the Jacobi identity and

\begin{equation}
C_{\alpha \beta }^{\alpha }=0
\end{equation}

Our previous definition of the persistence amplitude is inadequate because
gauge equivalent configurations are counted as distinct. To obtain a
satisfactory definition we must modify the measure of integration in such a
way that each gauge orbit is counted only once.

Let us introduce $linear$ functions $f^{\alpha }\left[ \Psi ^{i}\right] $
which are $not$ gauge invariant; in particular, we request

\begin{equation}
\Delta _{\beta }^{\alpha }=f_{,j}^{\alpha }Q_{\beta }^{j}
\end{equation}
to have an inverse. Then

\begin{equation}
\int D\varepsilon ^{\alpha }\;\mathrm{Det}\left[ \Delta _{\beta }^{\alpha
}\left[ \Psi ^{i}\left[ \varepsilon \right] \right] \right] \;\delta \left[
f^{\alpha }\left[ \Psi ^{i}\left[ \varepsilon \right] \right] -c^{\alpha
}\right] =1
\end{equation}
(provided the argument of the delta functions vanishes somewhere), and we
may write

\begin{equation}
Z\left[ 0\right] =\int D\varepsilon ^{\alpha }\;\int D\Psi ^{i}\;\mathrm{Det}%
\left[ \Delta _{\beta }^{\alpha }\left[ \Psi ^{i}\left[ \varepsilon \right]
\right] \right] \;\delta \left[ f^{\alpha }\left[ \Psi ^{i}\left[
\varepsilon \right] \right] -c^{\alpha }\right] \;e^{iS\left[ \Psi
^{i}\right] }  \label{zetauno}
\end{equation}

Since both the classical action and the volume element are gauge invariant,
this is the same as

\begin{equation}
Z\left[ 0\right] =\int D\varepsilon ^{\alpha }\;\int D\Psi ^{i}\;\mathrm{Det}%
\left[ \Delta _{\beta }^{\alpha }\left[ \Psi ^{i}\right] \right] \;\delta
\left[ f^{\alpha }\left[ \Psi ^{i}\right] -c^{\alpha }\right] \;e^{iS\left[
\Psi ^{i}\right] }  \label{zetados}
\end{equation}
This means the group volume may be factored out, thus defining

\begin{equation}
Z\left[ 0\right] =\int D\Psi ^{i}\;\mathrm{Det}\left[ \Delta _{\beta
}^{\alpha }\left[ \Psi ^{i}\right] \right] \;\delta \left[ f^{\alpha }\left[
\Psi ^{i}\right] -c^{\alpha }\right] \;e^{iS\left[ \Psi ^{i}\right] }
\label{zetaFP}
\end{equation}
In this expression, the correct integration measure is displayed.

It is interesting to check explicitly that this expression is independent of
the gauge fixing conditions $f^\alpha .$ Suppose we replace the functions $%
f^\alpha $ by new gauge fixing conditions $\mathbf{f}^\alpha =f^\alpha
+\delta f^\alpha .$ At the same time, we change variables in the functional
integral to new fields $\mathbf{\Psi }^i$ defined by the condition that $%
\mathbf{f}^\alpha \left[ \Psi \right] =f^\alpha \left[ \mathbf{\Psi }\right]
.$ This change is actually a gauge transform, since we may choose

\begin{equation}
\Psi ^i=\mathbf{\Psi }^i-Q_\alpha ^i\varepsilon ^\alpha
\end{equation}
where $\varepsilon ^\alpha =\left[ \Delta ^{-1}\right] _\beta ^\alpha \delta
f^\beta .$ Then the argument of the delta function as well as the exponent
remain unchanged. Since the gauge parameters may be field dependent, the
volume element is not invariant

\begin{equation}
D\Psi ^i=D\mathbf{\Psi }^i\;\mathrm{Det}\frac{\delta \Psi ^j}{\delta \mathbf{%
\Psi }^i}=D\mathbf{\Psi }^i\;\left\{ 1-\left( \varepsilon ^\gamma Q_\gamma
^j\right) _{,j}\right\}
\end{equation}

Concerning the functional determinant, we have

\begin{eqnarray}
\mathbf{\Delta }_\beta ^\alpha &=&\frac{\delta \mathbf{f}^\alpha \left[ \Psi
\right] }{\delta \Psi ^j}Q_\beta ^j\left[ \Psi \right] =\frac{\delta \mathbf{%
\Psi }^i}{\delta \Psi ^j}\frac{\delta \mathbf{f}^\alpha \left[ \Psi \right] 
}{\delta \mathbf{\Psi }^i}Q_\beta ^j\left[ \Psi \right] \\
&=&f_{,i}^\alpha Q_\beta ^i-f_{,i}^\alpha Q_{\beta ,j}^iQ_\gamma
^j\varepsilon ^\gamma +f_{,i}^\alpha \left[ Q_\gamma ^i\varepsilon ^\gamma
\right] _{,j}Q_\beta ^j \\
&=&f_{,i}^\alpha Q_\beta ^i-f_{,i}^\alpha Q_{\beta ,j}^iQ_\gamma
^j\varepsilon ^\gamma +f_{,i}^\alpha Q_{\gamma ,j}^i\varepsilon ^\gamma
Q_\beta ^j+f_{,i}^\alpha Q_\gamma ^i\varepsilon _{,j}^\gamma Q_\beta ^j \\
&=&\Delta _\gamma ^\alpha \left[ \delta _\beta ^\alpha +C_{\delta \beta
}^\gamma \varepsilon ^\delta +\varepsilon _{,j}^\gamma Q_\beta ^j\right]
\end{eqnarray}
Since $C_{\delta \gamma }^\gamma =0,$

\begin{equation}
D\Psi ^i\mathrm{Det}\left[ \mathbf{\Delta }_\beta ^\alpha \left[ \Psi
^i\right] \right] =D\mathbf{\Psi }^i\mathrm{Det}\left[ \Delta _\beta ^\alpha
\left[ \mathbf{\Psi }^i\right] \right].
\end{equation}
This concludes the proof of gauge independence.

Since the persistence amplitude is independent of the $c^\alpha ,$ we may
further average over them with a Gaussian weight $exp\left[ ic^\alpha
c_\alpha /2\sigma \right] $. Also the functional determinant may be
expressed as a path integral over Grassmann fields

\begin{equation}
\mathrm{Det}\left[ \Delta _{\beta }^{\alpha }\right] =\int D \zeta^{\alpha
}D\eta ^{\beta }\;exp\left\{ -\zeta _{\alpha }\Delta _{\beta }^{\alpha }\eta
^{\beta }\right\}
\end{equation}

The final expression looks like an ordinary persistence amplitude for a
theory with a modified action functional

\begin{equation}
Z\left[ 0\right] =\int D\Psi ^iD\zeta ^\alpha D\eta ^\beta
\;e^{iS_{FP}\left[ \Psi ^i,\zeta ^\alpha ,\eta ^\beta \right] }
\label{zetaFPG}
\end{equation}

\begin{equation}
S_{FP}=S\left[ \Psi ^{i}\right] +\frac{1}{2\sigma }f^{\alpha }f_{\alpha
}+i\zeta _{\alpha }\Delta _{\beta }^{\alpha }\eta ^{\beta }
\end{equation}

We may use this action to build an effective action in the usual way. If the
gauge fixing functions $f$ are chosen conveniently, it is possible to make
the effective action gauge invariant as well, thereby becoming the GIEA.

\subsection{The gauge invariant effective action \protect\cite{GIEA}}

We now belabor the dependence of the effective action on the gauge fixing
functions $f^\alpha .$ We have seen that the persistence amplitude is
independent of the choice of gauge fixing conditions. Since the effective
action reduces to the logarithm of the persistence amplitude when the
equations of motion hold (for short, ``on shell''), the variation of the
effective action with respect to the $f^\alpha $ must be related to the
first derivatives of the effective action with respect to the background
fields.

To make this explicit, let us consider a change in the gauge fixing
condition $f^\alpha \rightarrow f^\alpha +\delta f^\alpha .$ As we have
seen, the volume element, action, gauge fixing condition and ghost action
remain invariant if we simultaneously make a gauge transformation $\Psi
^i\rightarrow \Psi ^i-Q_\alpha ^i\varepsilon ^\alpha ,$ where $\varepsilon
^\alpha =\left[ \Delta ^{-1}\right] _\beta ^\alpha \delta f^\beta .$ But the
source terms are not invariant, whereby

\begin{equation}
\delta \Gamma =\Gamma _{,i}\left[ e^{-i\Gamma }\int D\Psi ^iD\zeta ^\alpha
D\eta ^\beta \;Q_\alpha ^i\varepsilon ^\alpha e^{i\left[ S_{FP}\left[ \Psi
^i,\zeta ^\alpha ,\eta ^\beta \right] -\Gamma _{,k}\left( \Psi ^k-\Phi
^k\right) \right] }\right]
\end{equation}

This formula is the key to define a gauge invariant effective action. The
idea is to allow the gauge fixing conditions $f^\alpha $ to depend
parametrically on the background fields $\Phi ^i,$ in such a way that the
total variation of the effective action is orthogonal to gauge trajectories

\begin{equation}
\frac{d\Gamma }{\delta \Phi ^i}Q_\alpha ^i\left[ \Phi \right] =\left[ \frac{%
\delta \Gamma }{\delta \Phi ^i}+\frac{\delta \Gamma }{\delta f^\beta }\frac{%
\delta f^\beta }{\delta \Phi ^i}\right] Q_\alpha ^i\left[ \Phi \right]
\equiv 0
\end{equation}
In view of the previous formula, this means

\begin{equation}
\Gamma _{,i}Q_\alpha ^j\left[ \Phi \right] e^{-i\Gamma }\int D\Psi ^iD\zeta
^\alpha D\eta ^\beta \;\left[ \delta _j^i+Q_\gamma ^i\left[ \Psi \right]
\left[ \Delta ^{-1}\right] _\beta ^\gamma \left[ \Psi \right] \frac{\delta
f^\beta }{\delta \Phi ^j}\right] e^{i\left[ S_{FP}\left[ \Psi ^i,\zeta
^\alpha ,\eta ^\beta \right] -\Gamma _{,k}\left( \Psi ^k-\Phi ^k\right)
\right] }\equiv 0
\end{equation}
Since the $Q_\alpha ^i$ are linear, we may also write

\begin{equation}
\Gamma _{,i}e^{-i\Gamma }\int D\Psi ^iD\zeta ^\alpha D\eta ^\beta \;Q_\gamma
^i\left[ \Psi \right] \left[ \Delta ^{-1}\right] _\beta ^\gamma \left[ \Psi
\right] \left[ \frac{\delta f^\beta }{\delta \Psi ^j}Q_\alpha ^j\left[ \Psi
\right] +\frac{\delta f^\beta }{\delta \Phi ^j}Q_\alpha ^j\left[ \Phi
\right] \right] e^{i\left[ S_{FP}\left[ \Psi ^i,\zeta ^\alpha ,\eta ^\beta
\right] -\Gamma _{,k}\left( \Psi ^k-\Phi ^k\right) \right] }\equiv 0
\end{equation}

The simplest way to satisfy this condition is to choose gauge fixing
conditions $f^{\alpha }$ invariant under a simultaneous transformation of $%
\Phi $ and $\Psi $ fields, in which case the brackets vanish identically.
Observe that since the total derivative of the GIEA is proportional to the
partial derivative (that is, with the gauge fixing condition kept fixed),
the equations of motion derived from one or the other are equivalent.

\end{document}